\documentclass[sigconf]{acmart}

\AtBeginDocument{%
  }

\acmConference[]{}{}{}

\usepackage{hyperref}

\begin{document}

\title{LLM-based Control Code Generation using Image Recognition}

\author{Heiko Koziolek}
\email{heiko.koziolek@de.abb.com}
\orcid{0000-0002-8805-6206}
\affiliation{%
  \institution{ABB Research}
  \country{Germany}
  }

\author{Anne Koziolek}
\email{koziolek@kit.edu}
\affiliation{%
  \institution{Karlsruhe Institute of Technology}
  \country{Germany}
}


\renewcommand{\shortauthors}{Koziolek et al.}

\begin{abstract}
LLM-based code generation could save significant manual efforts in industrial automation, where control engineers manually produce control logic for sophisticated production processes. Previous attempts in control logic code generation lacked methods to interpret schematic drawings from process engineers. Recent LLMs now combine image recognition, trained domain knowledge, and coding skills. We propose a novel LLM-based code generation method that generates IEC 61131-3 Structure Text control logic source code from Piping-and-Instrumentation Diagrams (P\&IDs) using image recognition. We have evaluated the method in three case study with industrial P\&IDs and provide first evidence on the feasibility of such a code generation besides experiences on image recognition glitches.
\end{abstract}


%

\maketitle

\section{Introduction}
Industrial process automation supports many production processes, e.g., for chemical plants, steel mills, or paper production and covers a 20 BUSD market. Real-time embedded automation controllers read vast amounts of sensor data in such processes, execute control logic, and write outputs to actuators, such as pumps, valves, or motors. Control engineers program these controllers using standardized programming notations, such as IEC 61131-3 Structured Text (ST), whose syntax was inspired by the Pascal programming language~\cite{Koziolek2020a}. Control programming is still a largely manual process and could yield significant cost savings from automated code generation.

Requirements for control programming are encoded as schematic CAD-drawings (i.e., Piping-and-Instrumentation Diagrams, P\&IDs), tables, and prose text by process engineers~\cite{Koziolek2020}. Control engineers manually interpret these requirements to design and implement control logic. The manual interpretation is a cognitive challenge due to the complexity of P\&IDs with hundreds of intricate instruments embedded into complex topological structures~\cite{Berlet2021}. Thus, this process is time-intensive, costly, and error-prone~\cite{Hollender2010}.

Due to the potential huge financial impact of code generation, researchers have proposed many approaches in the past~\cite{Koziolek2020}. Previous approaches involving P\&IDs relied on custom object-oriented notations (e.g.,~\cite{Drath2006,Haestbacka2011,Koziolek2020a}), while in practice still mainly rasterized diagrams are used due to convenience, missing standards, and IP risks. Previous approaches for image recognition on rasterized P\&IDs applied deep learning techniques and achieved high precision and recall (e.g.,~\cite{Kang2019,Yun2020,Kim2022}), but used limited training data sets and required still substantial manual rework. Large language models (LLM) have been found to generate IEC 61131-3 ST code well~\cite{Koziolek2023}, and have also been used for image recognition on hand-written sketches or screenshots to support code generation.

We propose a novel LLM-based control code generation method utilizing LLM-trained image recognition, domain knowledge, and code generation capabilities for industrial control logic. The method involves prompting an LLM with P\&ID images, asking it to recognize topological structures, and then iteratively generating control logic source code. Control engineers can feed the results into control logic programming environments, compile it, test it, and deploy it to automation controllers to then control complex production processes.

For the scope of this paper, we have tested the method in three exploratory case studies, applying it to P\&IDs from large process plants. We selected ChatGPT/GPT4V and prompted to generate IEC 61131-3 ST control logic. The code was fed into OpenPLC, an open-source programming environment, and checked for syntactical and functional correctness. We found mediocre image recognition capabilities with several glitches, but good code generation capabilities that depend on the prompt fidelity. Our method can improve with future LLMs and be automated to achieve high programming cost savings in a non-interactive process.

The next section provides background on control logic programming and introduces our P\&ID-based code generation method. Section 3 describes the first evaluation of the method generating ST-code for three heterogeneous P\&IDs and exploring image recognition and code generation capabilities. It also discusses threats to the validity of our approach. Section 4 reviews related work in different areas, before Section 5 concludes the paper.

\newpage

\section{LLM-based Control Code Generation}
\subsection{Background}
\begin{figure}[htbp]
  \includegraphics[width=\columnwidth]{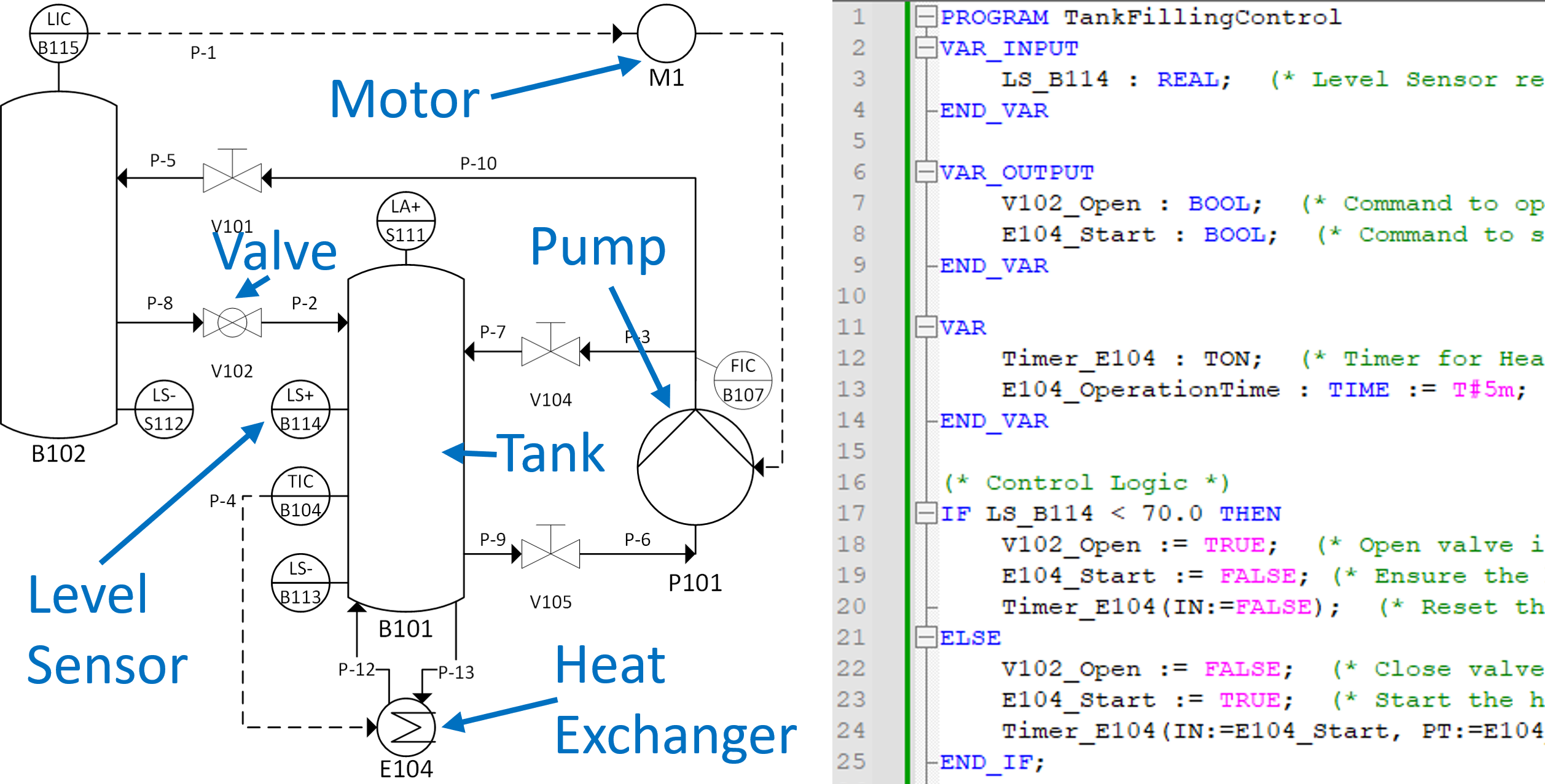}
  \caption{Piping and instrumentation diagram (P\&ID) and corresponding IEC 61131-3 Structured Text Control Logic to fill the tank B101 and heat its contents for 5 minutes}
  \label{fig:p-id-st}
\end{figure}

Many industrial automation processes, such as chemical process plants, material handling systems, or industrial drives and motors, are programmed using IEC 61131-3 \textbf{Structured Text} (ST). The language syntax was inspired by Pascal and C. For illustration, Fig.~\ref{fig:p-id-st} depicts a simple ST program on the right-hand side. ST provides typical variable declarations, control structures, operators, and functions. 
Unlike linear general-purpose programming execution models, ST-code is assigned to cyclicly executing tasks (e.g., every 100 ms) and continuously fed with new sensor data (e.g., new temperature or pressure values) as input for the control logic. Many commercial and open-source programming environments and execution runtimes are available. Previous experiments have shown that GPT4 handles ST-code generation well~\cite{Koziolek2023}. 

Control engineers use ST-code to express different types of \textbf{control strategies} in industrial automation. For example, PID (proportional, integral, derivative) control loops are used for maintaining a control variable (e.g., the filling level of a tank) at a desired set point (e.g., tank 70 percent full). Interlocks are safety-related mechanisms that link certain types of automation equipment together. For example, if a tank is filled more than 90 percent an interlock between a level sensor alarm and a feeding pump could ensure that the pump is automatically deactivated to prevent tank bursting. Another control strategy is sequential logic expressed as ST-code, which for example is used for start-up and shut-down procedures as well as for batch productions.

Process engineers usually specify the requirements for these control strategies, often supported by so-called \textbf{Piping and Instrumentation Diagrams} (P\&ID). Fig.~\ref{fig:p-id-st} shows a simplified example of a P\&ID on the left-hand side. It depicts piping, vessels, control valves, and sensors, among other process components. PID control loops may be directly specified in the diagram, e.g., LIC\_B115 in the top left corner of Fig.~\ref{fig:p-id-st} is a PID control loop for the level in tank B102. Interlocks can either be directly specified in the diagram or be derived by analyzing the process topology. Sequential logic for starting or stopping a complex production process with dozens of tanks and pumps may also be derived by interpreting P\&IDs. Process engineers or engineering contractors model most P\&IDs today with computer-aided design (CAD) tools (e.g., Autodesk P\&ID) and still distribute them as print-outs or rasterized PDFs, which complicates direct algorithmic processing without image recognition~\cite{Arroyo2016}.

Besides P\&IDs, process engineers also use \textbf{I/O lists} and \textbf{control narratives} to express automation requirements for control engineers. I/O lists are typically large tables where each entry represents an analog or digital signal associated with a particular sensor or actuator. The tables specify characteristics of the I/O signals and may already contain alarm limit thresholds and desired setpoints. Control narratives are prose text requirements specifications and express the desired control strategies using natural language. Usually, these texts are formulated independently of a particular control system and must be translated to vendor-specific programming notations and control function blocks.

\subsection{P\&ID-based Code Generation Method}
\begin{figure}[htbp]
  \includegraphics[width=\columnwidth]{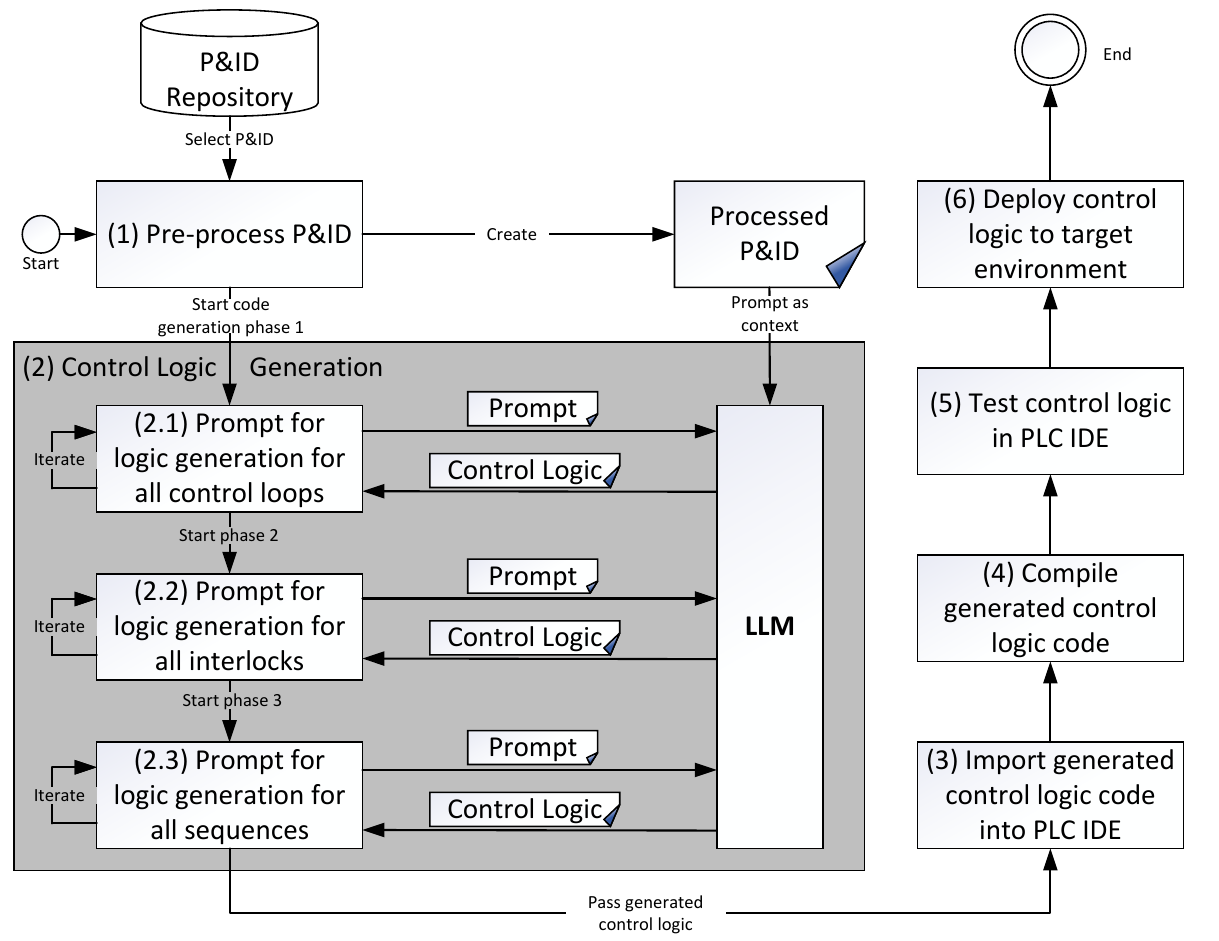}
  \caption{P\&ID-based Control Logic Code Generation Method: six steps from P\&ID to IEC 61131-3 ST}
  \label{fig:method}
\end{figure}

Our method aims at generating control logic by substituting the human P\&ID interpretation and human code writing to the image recognition and code generation capabilities of a large language model (LLM). The purpose of the method is to speed up the implementation of control logic and also to increase the control logic quality. While we have not automated the entire method yet, it should be possible to achieve a high level of automation and limit human interaction to supervision and testing in the future. This could save significant engineering costs and enable automation for processes where it is currently cost-prohibitive. The method is independent of a particular P\&ID notation, programming notation, or LLM. It consists of six steps depicted in Fig.~\ref{fig:method}.

For an automation project, the control engineers would receive several P\&IDs (among other documents) from an automation customer or intermediate engineering contractor. Large projects often include hundreds of complex A1-sized P\&IDs with hundreds of instruments each, each 
modeling individual process plant segments. The P\&IDs need to be pre-processed (\textbf{Step 1}) to make them processable by an LLM that supports image recognition. For example, P\&IDs provided on paper need to be scanned to be available as digital images. The images may need color and contrast adjustments to improve image recognition quality. Furthermore, the used LLM may be limited by the amount of information to process at once. In this case, complex P\&IDs need to be segmented (possibly automatically), so that smaller image cutouts can be prompted to the LLM

\textbf{Step 2} contains the actual code generation and is divided into three sub-steps. In Step 2.1 first prompts for generating control logic for control loops are issued to the LLM. This can be based on first prompting the LLM to detect all explicitly specified control loops in the current P\&ID image. Each control loop may be assigned a pre-defined PID function block in a batch run. Step 2.1 may be executed iteratively until all elements of the current P\&ID are processed. In Step 2.2 the LLM is prompted to recognize required interlocks in the current P\&ID and then generate ST-code for them. Finally, in Step 2.3 the LLM is prompted for sequential logic, e.g., start-up and shutdown procedures for the shown process. 

It is conceivable to execute the prompting and collecting of control logic from the LLM in a non-interactive batch run by a software agent similar to AI agents, such as AutoGPT\cite{Hadi2023}. Such a batch run may also include quality checks and intermediate checks, in addition to automatically generated follow-up prompts. Whether full automation is feasible, however, still needs to be researched in future work.

After generating the control logic in Step 2, in \textbf{Step 3} the control engineer or a software agent imports the LLM-generated code into a control logic integrated development environment (IDE). Such an IDE allows human reviews of the code in language-specific editors that support syntax highlighting, code collapsing, or auto-completion. In \textbf{Step 4}, the code can be compiled, e.g., to C-code or directly to machine code. Afterwards, in \textbf{Step 5}, the user can test and debug the code in a simulation environment in the IDE. Finally, if the code fulfills the function and non-functional requirements as verified by simulation runs, in \textbf{Step 6} the compiled code can be deployed to the target industrial controllers, e.g., real-time embedded micro-controllers or Industrial PCs. After all the required automation equipment is installed on-site, the code can be started to control the production process.

The entire method is generic and allows for many possible extensions and refinements. Besides P\&IDs other control strategy requirements (e.g., I/O lists and control narratives) could be fed to the LLM for more context. Besides IEC 61131-3 ST, the method could generate other typical control logic notations, such as function block diagrams or sequential function charts. Instead of PDF-based images of P\&IDs, so-called smart P\&IDs based on object-oriented notations could be fed into the process as text documents to avoid complications from possibly unreliable image recognition. Besides the core control logic, testing and simulation code could be generated with the method if appropriate prompts can be formulated. Low-level prompts could be replaced by formulating higher-level objectives so that the LLM itself finds an optimal procedure to achieve them. However, for the scope of this paper, we restrict the following evaluation of the method to a first exploratory test of the core concepts.

\newpage
\section{Evaluation}
\subsection{Methodology}
For evaluation, we tested core concepts of the method on concrete P\&IDs from industrial projects. We chose the generation of IEC 61131-3 ST code in favor of other notations. As LLM, we used GPT4 in a version of November 2023. Following an exploratory approach, we conducted interactive sessions using the ChatGPT chat interface and did not attempt to run batch queries through the API. This allowed us to react to the answers and adapt subsequent prompts to reveal more insights about the particular cases.

To avoid vendor bias and improve replicability, we chose an open-source IEC 61131-3 programming environment, namely OpenPLC~\cite{Alves2014}. It includes the IEC2C compiler to translate the ST-code to ANSI-C and then compile it to machine code for the OpenPLC IEC 61131-3 runtime. We used a separately available Python tool called OpenPLC-Importer to feed the generated ST-code into the OpenPLC Editor. We also used its integrated simulator to perform test runs of the control logic.

We performed the code generation on three different P\&IDs. Many P\&IDs from customer projects contain proprietary information (e.g., recipes, procedures). To avoid disclosing protected intellectual property, we used publicly available P\&IDs from industrial cases. The P\&IDs are either directly from real process plants or created as exemplars by industry consortia. They use different kinds of symbols, which allows us to check GPT4's robustness against different notations. Previous approaches to perform image recognition on P\&IDs were often trained for a specific kind of P\&ID notation.

Our evaluation was exploratory and focused on the use case of code generation. A systematic evaluation of an LLM's image recognition ability should use precision and recall metrics for certain shapes, pipes, or symbols, as done by Kim et al.~\cite{Kim2022}. We restricted our analysis to representative samples for specific instruments and performed no batch runs. We also did not have the original control logic available that was used in the production plants built after the P\&IDs. Therefore, without a ground truth, we only performed manual syntax and plausibility checks. 

We did not compile all the generated code, since the syntactic correctness of ST-code generation using GPT4 has been demonstrated in other works (e.g. \cite{Koziolek2023}). We prompted ChatGPT to generate comparably simple and sometimes abstract code since we lacked context information about the P\&IDs and also did not want to introduce vendor bias for example by using proprietary ST-code function block libraries. The LLM was used as-is, without any retrieval-augmented generation or fine-tuning.

Our chat sessions for each P\&ID were included in a single conversation each, so that the context of previous prompts and answers for the same P\&ID may have affected the outcome of later queries. Each session roughly followed steps 2.1 to 2.3 of our methods, but we introduced smaller case-specific adaptations to explore specific aspects. We only performed a few repetitions of individual prompts to improve the results. We used a few prompt engineering techniques, e.g., to generate self-contained code or to generate specific comment notations supported by OpenPLC. Full logs of our chat sessions are available in supplemental online material (\url{https://zenodo.org/records/10148136}).

\newpage
\subsection{Case Study 1: Eastman Chemical}
\begin{figure}[htbp]
  \includegraphics[width=\columnwidth]{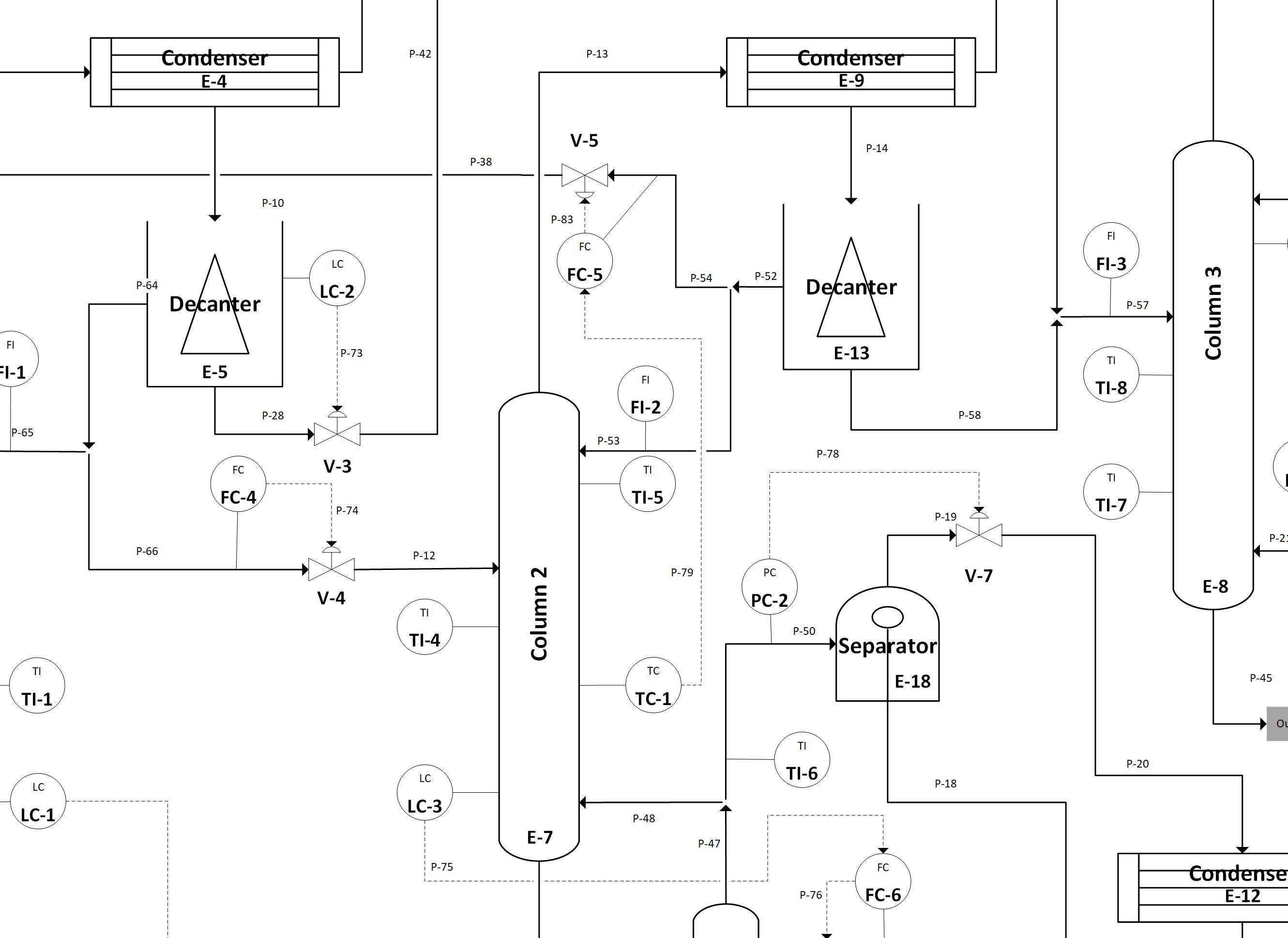}
  \caption{Eastman Chemical Plant P\&ID (cutout)}
  \label{fig:eastman1}
\end{figure}

A process plant of Eastman Chemical Company, US, was previously subject to a plant-wide disturbance analysis by Thornhill et al.~\cite{Thornhill2003}. The available P\&ID drawing (Fig.~\ref{fig:eastman1}) contains three distillation columns, two decanters, and several recycle streams~\cite{Berlet2021}. The specified process also includes 14 controlled actuators and 15 indicators, e.g., for temperature, pressure, and level. 

In the P\&ID, small circles with tagnames including the letter 'C' indicate the control points. Upon our prompt, ChatGPT was able to correctly recognize all 14 control points, i.e. 7 for flow, 3 for level, 2 for pressure, and 2 for temperature.  Algorithms could assign PID function blocks to those controllers that directly control an individual actuator based on a single sensor reading. 

We prompted ChatGPT to identify feedforward cascading control schemes, which refer to situations where the control output of a primary controller (e.g., for temperature) serves as input for a secondary controller (e.g., for flow). The identification of such schemes requires finding topologically connected controllers (i.e., circles connected by dashed lines). ChatGPT reported four such control schemes in the P\&ID, of which only one was correct. ChatGPT did not recognize four other such control schemes in the diagram, and we prompted to correct its answer (e.g., TC-1 is in a feedforward control scheme with FC-5, see Fig.~\ref{fig:eastman1}). 

ChatGPT then generated 46 lines of IEC 61131-3 ST code for TC-1 and FC-5 based on the following prompt: ``Write a self-contained IEC 61131-3 ST function block for the feedforward cascading loop including TC-1 and FC-5. Assume both controllers are PID controllers, provide plausible PID parameters. Name the input variables after the connected sensors and declare them as input variables. Name the output variables after the attached controllers or valves and declare them as output variables. For comments in the source code only use the (* … *) notation, not //.''

\begin{figure}[htbp]
  \includegraphics[width=\columnwidth]{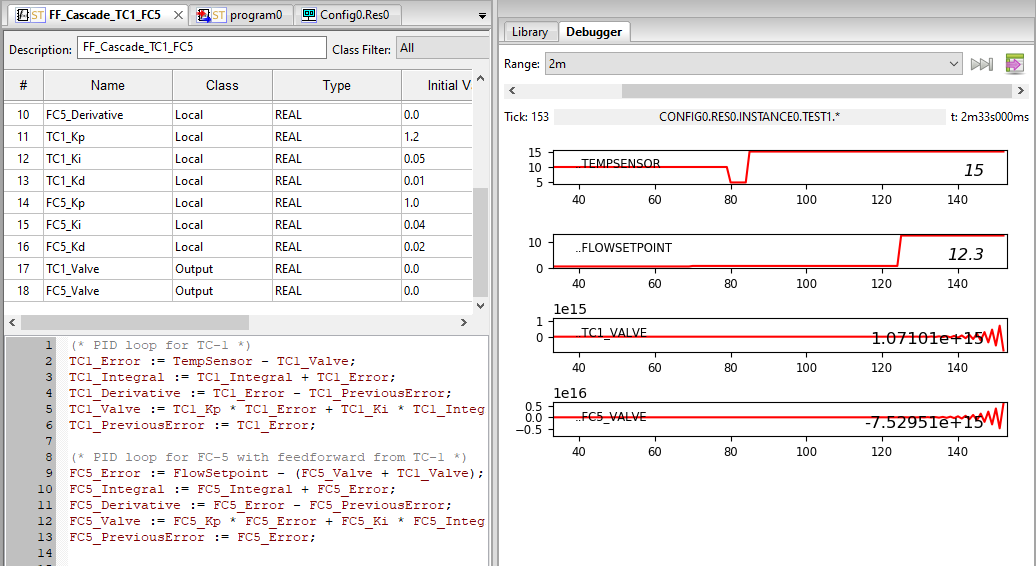}
  \caption{Generated ST source code and simulation test run results in OpenPLC for a feedforward cascading control scheme}
  \label{fig:sim1}
\end{figure}

The OpenPLC Import Tool allowed to add the function block into an OpenPLC project (Fig.~\ref{fig:sim1}). After instantiating the function block in a program, the code successfully compiled. This demonstrated that the generated code is syntactically compliant with IEC 61131-3. For debugging, we ran a simulation for several minutes in OpenPLC and manually manipulated the temperature values. This demonstrated that the code fulfills the minimal functional specifications and can now be fine-tuned for the specific control scheme. PID parameters and set point values need to be adjusted to the chemical process, but these were not included in the P\&ID. 

Afterward, we asked ChatGPT to provide the interlocks required for distillation column E-7 (cf. Fig.~\ref{fig:eastman1}). ChatGPT's answer correctly specified the required level and temperature interlocks and the required effects, although it omitted concrete tag names. ChatGPT also stated required high and low pressure interlocks, despite absent pressure sensors for E-7 in the P\&ID (Fig.~\ref{fig:eastman1}). Futhermore, it generated a reboiler flow interlock and an emergency shutdown interlock for E-7, both of which are plausible. Answering a follow-up prompt, ChatGPT then generated 76 lines of correct ST-code for the interlocks.

Finally, we asked ChatGPT to generate a startup sequence for the process, which requires understanding the flow in the piping structure. ChatGPT correctly identified the starting point of the shown process (i.e., In\_2\_Feed and valve V-1) and described a procedure that followed through reboiler E-19, column 1, decanter E-5, and column 2 in nine separate steps. We then prompted for ST code generation for step 2 of this procedure, i.e., startup of column 1. ChatGPT generated another 55 lines of ST for this step. The code included setpoints provided by us in the prompt, and also generated feed level and heat input adjustments. The logic was simple but adhered to the prompt and the P\&ID.

Despite several challenges in recognizing more complex topological structures, generating syntactically and functional ST-code from the P\&ID, in this case, was mostly successful. In further tests, we noticed ChatGPT failing to associate pipe labels to the respective pipes or hallucinating control points that were not included in the P\&ID. The currently generated code is still abstract, but can be detailed. In a real setting, ChatGPT would need more contextual information (e.g., set points, alarm limits, equipment dimensions). However, as demonstrated, using image recognition ChatGPT can utilize topological information encoded in a P\&ID, e.g., for cascading loops or startup sequences. Automating our manually executed prompts in a batch run (e.g., for all controllers, for all interlocks) would be straightforward.

\subsection{Case Study 2: DEXPI}
\begin{figure}[htbp]
  \includegraphics[width=\columnwidth]{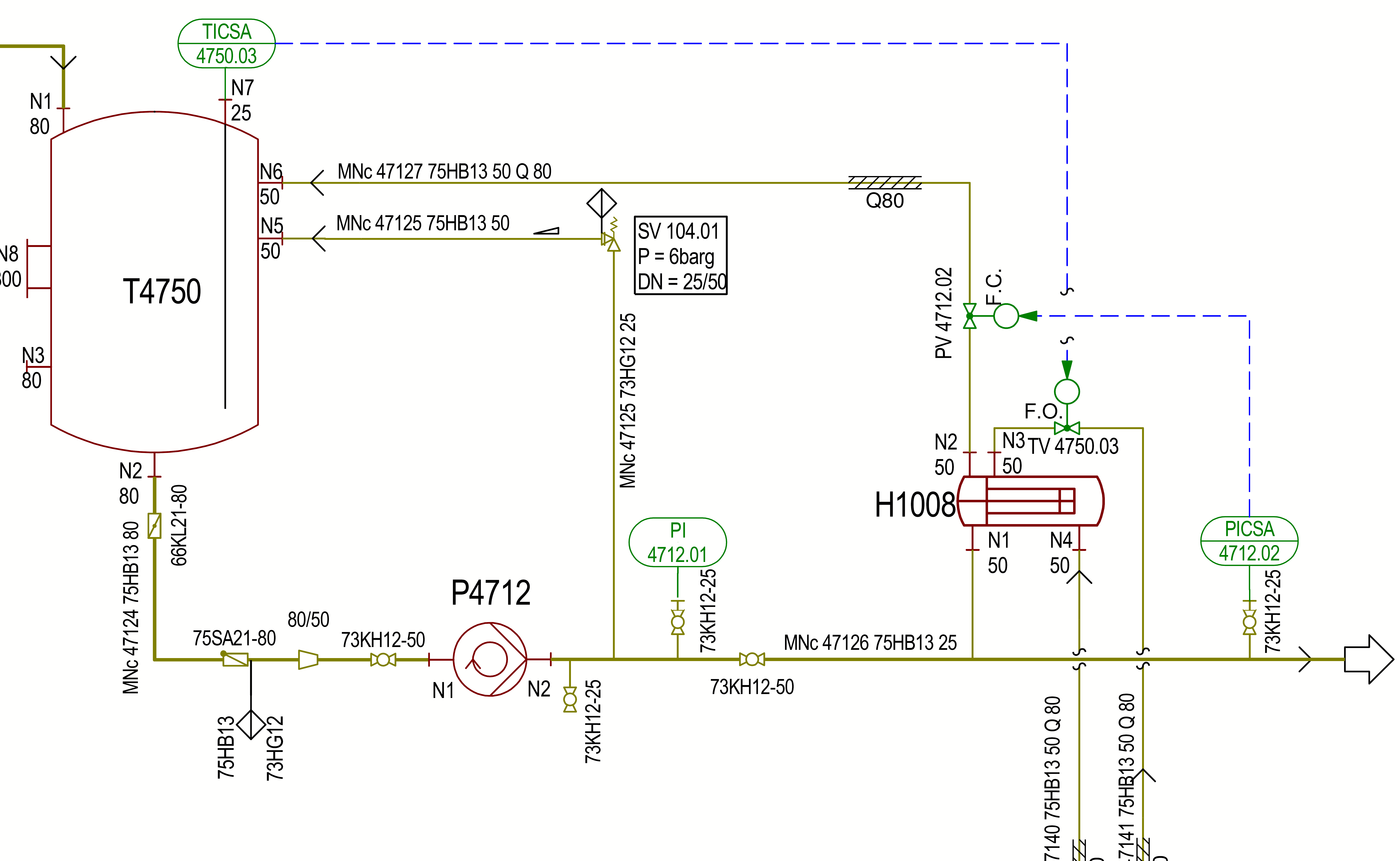}
  \caption{DEXPI Reference P\&ID (cutout)}
  \label{fig:dexpi1}
\end{figure}

DEXPI stands for ``Data Exchange in the Process Industry'' and has recently become a registered association for developing and promoting common data standards for chemical process plants. Large chemical companies, such as BASF, Equinor, or Shell, are members of this initiative, as well as vendors of popular CAD applications. The organization has specified an exemplary P\&ID for testing data exchange (Fig.~\ref{fig:dexpi1}). The diagram uses the ISO 10628 notation for P\&IDs and includes detailed pipe and equipment nozzle labels, as well as a few tables with equipment design parameters (e.g., maximum temperature for a tank or dimensions). The process contains one large tank, two pumps, two heat exchanges, and four instruments. It does not depict a real process, but was condensed to include many P\&ID elements to test DEXPI importers and exporters.

We first prompted ChatGPT for controllers in the P\&ID, which are here depicted with green ovals. ChatGPT correctly identified the temperature controller TICSA 4750.03 and the pressure controller PICSA 4712.02. However, it also included hand switch HS 4750.01 and pressure indicator PI 4712.01 in the list, which are not controllers. The P\&ID used the same graphical depiction with a green oval connected to a valve or pipeline.

We then asked for ST-code for PID controller PICSA 4712.02, for which ChatGPT generated 38 lines of code. While the PID control code was simple and correct, the generated input and output signal references in this case were all wrong. It generated an input variable called PV\_4712\_01, which is not visible in the P\&ID, but however plausible. For the output variable, it used a tag name for the flow orifice 4750.03, which is in proximity of PICSA 4712.02, but not connected to it. Instead, the correct output variable would have been PV 4712.02 (blue dashed line in the P\&ID). In this case, ChatGPT thus severely misinterpreted the P\&ID due to an erroneous image recognition or model inference. The blue connecting line in the diagram intersects with another dashed blue line belonging to TICSA 4750.03 and TV 4750.03. Human interpretation unlikely would have made this mistake.

Asked for interlocks for tank T4750 (Fig.~\ref{fig:dexpi1}), ChatGPT noted that it lacked specific information about the purpose and functionality of T4750, but tried to generate a generic list of plausible interlocks given the graphical specification. This list included interlocks for level, pressure, and temperature, as well as for pump protection, mixer/agitator protection, and emergency shutdown. We then decided to provide more information to ChatGPT and prompted for interlock ST-code generation, but included the tank dimension and design pressure, which were given as tables in the P\&ID into the prompt. 

\begin{figure}[htbp]
  \includegraphics[width=\columnwidth]{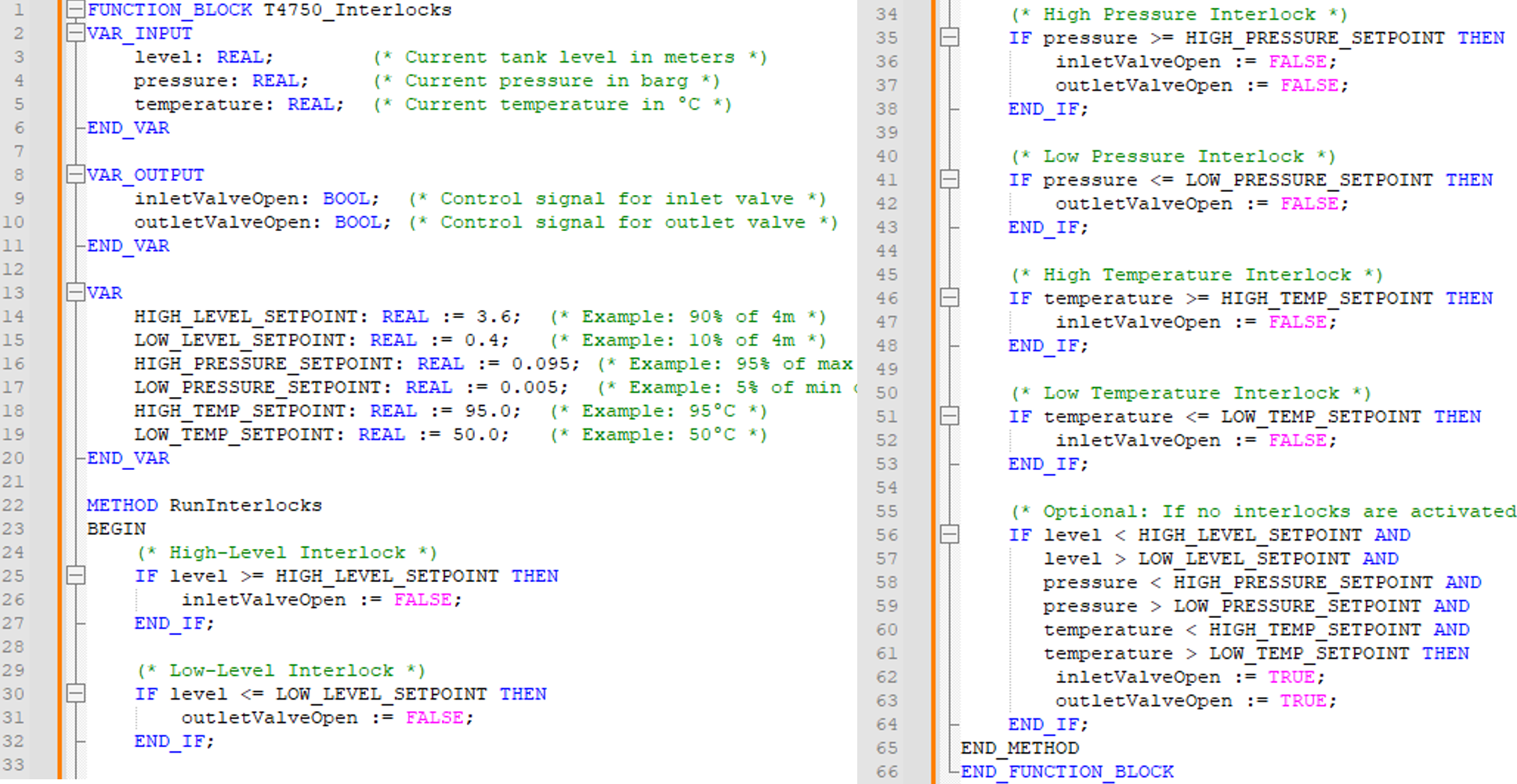}
  \caption{Generated ST code for interlocks includes plausibly generated alarm limits}
  \label{fig:dexpi-interlocks}
\end{figure}

ChatGPT generated 66 lines of ST-code (Fig.~\ref{fig:dexpi-interlocks}) for the interlocks and included the additional specifications in an informed manner. For example, it plausibly defined a high-alarm limit at 90 percent of the 4 m tank height, which was 3.6 meters. The generated code was still abstract and did not refer to concrete tag names. It included commands to inlet and outlet valves which are not directly visible in the P\&ID. While the code was a decent approximation of the target source code, ChatGPT correctly stated "Adjustments might be needed based on the real-world requirements". 

Finally, we prompted to generate a startup sequence. ChatGPT correctly identified that pump P4711 initially feeds the process and that P4712 needs to be started subsequently. 
ChatGPT then stated to activate the heat exchanger at 70 percent nominal capacity, which is plausible. Specifically for the tank filling of T4750, we prompted for ST-code generation, which was correctly executed and answered again with rather abstract code lacking concrete tagnames.

Further tests showed that ChatGPT had trouble finding the pipeline between tank T4750 and pump P4712. The cause could be that the path leading to the pump contains several pipe adapters and annotations and also does not include an (actually mandatory) arrowhead specifying the flow direction. For the same tank, we prompted for a list of the attached 7 nozzles, N1, N2, N3, N5, N6, N7, and N8. Interestingly, ChatGPT created a list of 8 nozzles, N1 to N8, including a non-existing N4. Here the statistics or model inference seem to have overtaken the image recognition when performing the text completion.

While the code generation in this case showed potential, none of the generated code snippets would be directly usable in practice. 
However, our prompts were still rather abstract and relied mostly only on the graphical information in the P\&ID. More contextual information in the prompts could lead to more practical code.

\subsection{Case Study 3: Butane Regeneration}
The third case tackles generating control logic for a butane regeneration air and water cooling system. A Korean engineering company provided the P\&ID drawing, which was used in 2019 for a study on symbol and text recognition for P\&IDs based on template matching~\cite{Kang2019}. Butane is a gas often used as fuel for portable lighters and the manufacturing of a wide range of chemicals. The specified process contained two water coolers, one air cooler with two fans, and a lot of instrumentation mostly for temperature and pressure.

\begin{figure}[htbp]
  \includegraphics[width=\columnwidth]{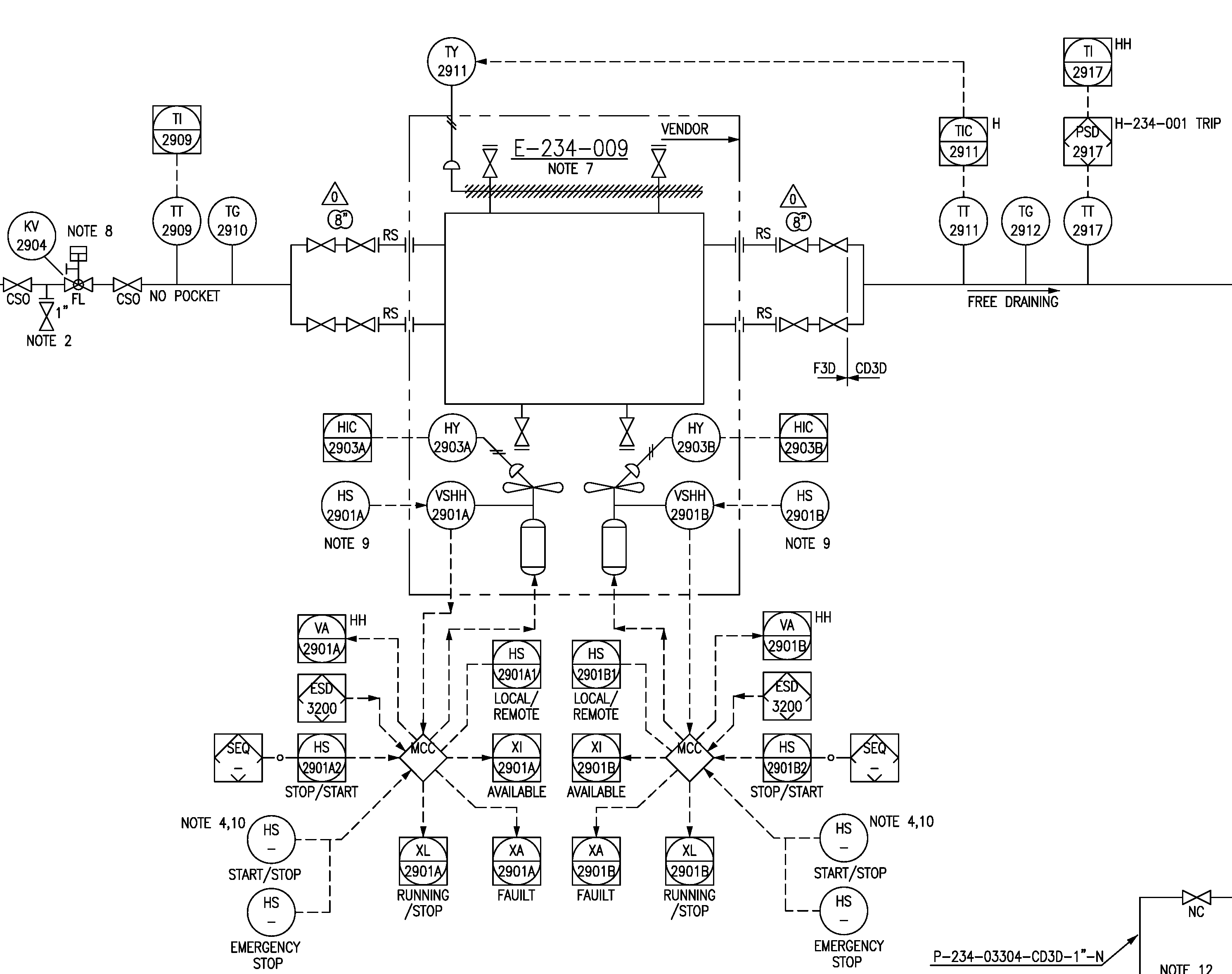}
  \caption{Butane regeneration air cooling P\&ID (cutout)}
  \label{fig:butane}
\end{figure}

When we first prompted ChatGPT to identify the controllers in the entire P\&ID, we received a list with several hallucinated controllers that did not appear in the image. We thus simplified the task and used only the cutout shown in Fig.~\ref{fig:butane} for image recognition tasks. This simplification may be justified, as only rare topological dependencies span the entire diagram. ChatGPT provided a shorter list of controllers for the P\&ID cutout, which however still contained temperature transmitters which are no controllers, as well as a few hallucinated controllers. ChatGPT correctly pointed to E-234-009 as an important vessel, but identified the two air fans as pumps.

Generating a PID control loop for the temperature controller TIC2911 yielded syntactically correct ST code, although ChatGPT did not use the correct tag names for input and output. ChatGPT was also able to generate a long list of 13 different interlocks for the E-234-009 vessel, including fan operation interlocks. Due to missing sensors, such interlocks would be impossible to implement. However, ChatGPT could generate appropriate ST-code for the interlocks including plausible min/max values for process parameters.

When we prompted ChatGPT for a detailed startup sequence for the air cooling system, it generated a 10-step procedure, which correctly included establishing the inflow, starting the air cooler, and starting the control loops. We then prompted ChatGPT to generate ST-code for the previously defined startup sequence, which yielded rather generic code that simply checked the inlet pressure and outlet temperature without any timings. 

\begin{figure}[htbp]
  \includegraphics[width=\columnwidth]{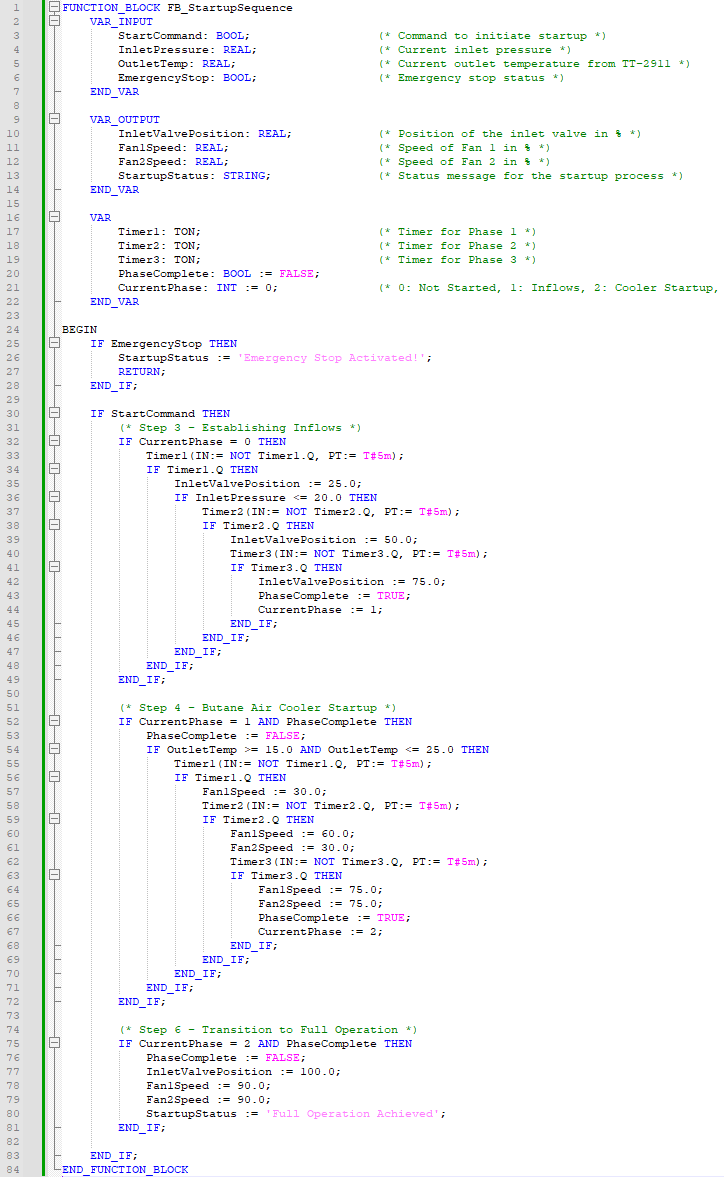}
  \caption{Generated ST-code for starting the butane regeneration air cooling system with timer blocks and state machine.}
  \label{fig:butane-startup}
\end{figure}

Therefore we refined the prompt for defining the startup sequence `` [...] Include valid analog ranges for the inlet valves. Provide concrete operational flow rates to target during startup. [...] Provide timings for gradually increasing fan speed and flow.''

This answer included concrete values for valve openings, as well as plausible timings for opening them. We used this to re-generate the startup ST-code. ChatGPT now generated 77 lines of ST-code with the concrete parameter values and matching timer function blocks. The startup process was partitioned into three phases. However, the code contained a bug starting phases 1 and 2 in parallel.

In a follow-up prompt, we pointed out the error to ChatGPT, which corrected the ST-code accordingly to implement a valid state machine. The result was 83 lines of ST-code, depicted in Fig.~\ref{fig:butane-startup}. We learned from this case that partitioning the P\&ID and providing more concrete prompts can improve the code.

\subsection{Threats to Validity}
We review the internal, external, and construct validity of our study. The \textbf{internal validity} refers to the extent to which a study can establish a causal relationship between manipulations of the independent variables leading to changes in dependent variables. A threat to the internal validity of our study is the inherent non-deterministic nature of LLMs leading to different answers for the same prompts. To counter this, we publish at least our precise prompts and the received answers as supplemental online material (\url{https://zenodo.org/records/10148136}). Researchers can reuse the prompts for replication studies. The perceived mediocre image recognition performance of GPT4 in our three cases should be similar in replication studies. Also, the speed of generating the ST-code (around 10 seconds) is not affected by our experimental settings.

The \textbf{construct validity} refers to the extent to which the tests actually measure what they claim to be measuring. In our case, we selected ChatGPT and GPT4 as typical constructs for an LLM, although in an eventual realistic large-scale code generation approach rather the LLM API than the chat interface would be used. We selected IEC 61131-3 ST as a typical construct for a control code programming notation, which is based on a widespread international standard. Furthermore, in previous experiments, GPT4 showed good syntactical knowledge of IEC 61131-3 ST and overall good code generation quality. We used PDF-based P\&IDs as constructs for process engineering requirements specifications, but not newer smart P\&IDs based on object-oriented notations, such as DEXPI/ISO15296. PDF-based P\&IDs are still most widespread in practice and cover more than 95\% of the existing plants.

We used P\&IDs from industrial settings, but in some cases processed cutouts of them. We ran typical control generation prompts. Although these do not cover all types of control logic, they should be representative. The use case of generating control logic source code from (legacy) P\&IDs may be artificial, since in practice for most projects such code is already available and does not need to be re-generated. Newer projects could start with smart P\&IDs that do not require image recognition for code generation. We argue however that still many projects use PDF-based P\&IDs, that code generation could be also done for other purposes (e.g., simulation or test code), and that many projects require additional control logic over the lifetime due to maintenance. We generated rather low-level ST-code and did not include library function blocks, which are often available in practice. 

The \textbf{external validity} refers to the extent to which a study's finding can be generalized to other contexts and settings. We argue that GPT4 can perform image recognition on a vast range of different P\&ID notations, although its exact training data is unknown. Furthermore, our code generation approach is not specific for a single subdomain of industrial automation, such as oil\&gas processing or pulp\&paper handling. ST-code can be used in all kinds of subdomains of industrial automation, and GPT4's domain knowledge also covers vastly different settings. The approach as such should also be transferable to other code notations, either other industrial coding notations, such as function block diagrams or sequential function charts, or other general purpose coding notations, such as C\#, C++, Python, or Java.

\section{Related Work}
We review 1) methods to generate control logic from P\&IDs, 2) methods to perform image recognition on P\&IDs, and 3) code generation using image recognition in other domains.

Koziolek et al.\cite{Koziolek2020} have surveyed several methods to \textbf{generate control logic from P\&ID} drawings. Drath et al.~\cite{Drath2006} use P\&IDs encoded as XML, apply a set of topological rules, generate an interlocking table, and then IEC 61131-3 ST. The AUKOTON tool~\cite{Haestbacka2011} maps XML-based P\&IDs and I/O lists into domain-specific models, before creating IEC 61131-3 ST using a PLCOpen generator. Thramboulidis et al.~\cite{Thramboulidis2011} derive SysML models from XML-encoded P\&IDs, which in turn can be transformed into PLCOpenXML control logic. Arroyo et al.~\cite{Arroyo2016} perform image recognition on rasterized P\&IDs as PDF files and synthesize low-fidelity simulation source code. Koziolek et al.~\cite{Koziolek2020a} derived IEC 61131-3 ST from object-oriented, smart P\&IDs using a rule engine. None of these works utilized LLMs.

Kim et al.~\cite{Kim2022} provide a recent overview of \textbf{methods to perform image recognition on P\&IDs} using deep learning and other approaches. For example, Kang et al.~\cite{Kang2019} use template matching and OCR to detect symbols, lines, and text in P\&IDs. Yu et al.~\cite{Yu2019} employ connectionist text proposal networks (CTPN) to perform symbol, text, and line recognition on P\&IDs and achieve a 91.6 percent accuracy for symbols. Yun et al.\cite{Yun2020} apply region-based convolutional neural networks on P\&IDs and find a 98\%  symbol recognition rate in their experiments. Kim et al.~\cite{Kim2021} perform deep learning and object character recognition on a data set of P\&IDs and report a 97.2\% precision for symbol recognition. A study by Kim at al.~\cite{Kim2022} uses deep neural networks on P\&IDs for symbol, text, and line recognition, reports an average precision of 99.5\% for topology reconstruction, but also finds that 2-4 hours of manual re-work are required for each P\&ID. None of these approaches utilized LLMs or generated IEC 61131-3 ST control logic.

Several other works tackle \textbf{code generation using image recognition in other application domains}. Karasneh et al.~\cite{Karasneh2013} perform image recognition on rasterized UML-diagrams and generated XMI-based files that could feed classical UML-based code generation tools. Chen et al.~\cite{Chen2022} conduct image recognition on 80 UML images as a precursor for code generation. Asiroglu et al.\cite{Asiroglu2019} recognize hand-drawn mock-ups for web pages and generate HTML code. Similarly, Yashaswini et al.\cite{Yashaswini2022} propose an HTML code generator for screenshot images and hand-drawn sketches. Camara et al.~\cite{Camara2023} found that ChatGPT had limitations regarding software modeling, citing syntactic and semantic deficiencies, and a lack of consistency and scalability. Another interesting direction is generating P\&IDs with LLMs, as performed by Hirthreiter et al.~\cite{Hirtreiter2023}. None of the other approaches for code generation used as sophisticated images as P\&IDs or generated control logic.

Compared to related work, our method combines the recent advancements of LLMs with previous work on code generation for industrial application cases~\cite{Koziolek2023}. Using pre-trained LLMs avoids custom LLM training and can be more flexible due to their large training sets. Unlike custom-trained models, it can also integrate the domain knowledge and code generation capabilities of LLMs into the code generation process. 

\section{Conclusions}
This paper has proposed a novel LLM-based code generation method specifically for control logic in industrial automation. We have evaluated the method by feeding P\&IDs to GPT-4V and testing its capabilities for recognizing topological structures and synthesizing code based on domain knowledge. While in its current version, the image recognition still showed several glitches, we provided evidence for the method's principle feasibility. Working code for complex automation tasks was generated within several seconds. Due to the structured requirements specifications, it is conceivable to largely automate the code generation process in this domain, thus advancing over typical interactive co-pilot code generators used in other domains.

Practitioners can already adopt the method in their projects and use our prompts as templates for formulating their own based on their specific use cases. Implementing tool support for the method and increasing its level of automation is an obvious next step that could be supported by developers. Practitioners could contribute to comprehensive and representative P\&IDs data sets that could be used to systematically analyze the image recognition capabilities of future LLMs. With our method, researchers get a starting point for combining research on deep learning-supported image recognition on P\&IDs with research on industrial code generation. Other researchers can refine the method by introducing additional forms of code generation or designing methods to perform automatic plausibility checks on the LLM outputs.

Future work involves testing the method on larger data sets, refining the used prompts for code generation, and adding more automation. In non-interactive batch runs, an entire P\&ID could be traversed with numerous prompts to generate a large collection of source code files. This approach could be extended to entire sets of P\&IDs for a large automation project. Such an approach would need sophisticated and partially automated means to check the correctness, plausibility, and compatibility of the generated code. 

Besides P\&IDs, other artifacts, such as I/O lists or control narratives should be fed into the code generation process to improve the code generation fidelity and correctness. This could be implemented using retrieval-augmented generation in a multi-modal prompting scheme. In the same manner, vendor-specific programming notations or pre-existing control function blocks could be fed into the generation to reduce manual re-works even further. Besides control logic, other kinds of code could generated, for example, simulation code or code for human machine interfaces.

\bibliographystyle{ACM-Reference-Format}
\bibliography{LLM4Code}

\end{document}